\documentclass[twocolumn,showpacs,preprintnumbers,amsmath,amssymb,aps]{revtex4-1}
\usepackage{amssymb,color,array}
\usepackage{graphicx}
\usepackage{amsmath}
\usepackage{hyperref}
\definecolor{rein}{rgb}{0,0,1}

\begin{document}

\title{Giant shot noise from Majorana zero modes in topological trijunctions}

\author{T. Jonckheere,$^1$ J. Rech,$^1$ A. Zazunov,$^2$ R. Egger,$^2$ A. Levy Yeyati,$^3$ and T. Martin$^1$} 
\affiliation{${}^1$~Aix Marseille Univ, Universit\'e de Toulon, CNRS, CPT, Marseille, France\\
${}^2$~Institut f\"ur Theoretische Physik, Heinrich Heine Universit\"at, D-40225 D\"usseldorf, Germany\\
${}^3$~Departamento de F{\'i}sica Te{\'o}rica de la Materia Condensada C-V, Condensed Matter Physics Center (IFIMAC) and Instituto Nicol{\'a}s Cabrera, Universidad Aut{\'o}noma de Madrid, E-28049 Madrid, Spain}

\date{\today}

\begin{abstract}
The clear-cut experimental identification of Majorana bound states in transport measurements still poses experimental challenges. We here show that the zero-energy Majorana state formed at a junction of three topological superconductor wires is directly responsible for giant shot noise amplitudes, in particular at low voltages and for small contact transparency. 
The only intrinsic noise limitation comes from the current-induced dephasing rate due to multiple Andreev reflection processes. 
\end{abstract}

\maketitle

\emph{Introduction.---}Majorana fermions have emerged as quasi-particles of central importance in modern condensed matter physics, e.g., for  topological superconductors (TSs) and in exotic phases with intrinsic topological order \cite{Nayak2008,Alicea2012,Leijnse2012,Beenakker2013,Sarma2015,Aguado2017,Lutchyn2018}.  In one-dimensional TS wires, spatially localized 
Majorana bound states (MBSs) are formed at the wire boundaries. The corresponding Majorana operator
represents a quasi-particle that equals its own antiparticle.  MBSs are associated with non-Abelian braiding statistics, and a pair of well-separated MBSs defines a non-local zero-energy fermion state.  Apart from the obvious fundamental interest,
 stable and robust realizations of zero-energy MBSs would also enable powerful topologically protected quantum information processing schemes \cite{Kitaev2001,Nayak2008,Sarma2015,Alicea2011,Plugge2017,Karzig2017}.    
 Over the past few years, many  experiments have reported evidence for MBSs either through the observation of conductance peaks in transport spectroscopy (with normal probe leads tunnel-coupled to MBSs) \cite{Mourik2012,Yazdani2014,Ruby2015,Albrecht2016,Deng2016,Nichele2017,Suominen2017,Gazi2017,Zhang2018,Vaiti2018} or from signatures of the $4\pi$ periodic Josephson current-phase relation in TS-TS junctions \cite{Deacon2017,Bocquillon2017,Laroche2018,Fornieri2018}.  
However, in principle both types of experiments are not able to firmly rule out alternative physical mechanisms. In fact, zero-bias anomalies are ubiquitous and could arise from many sources, e.g.,  subgap Andreev states \cite{Moore2018,Vuik2018} or disorder \cite{Altland2012,Liu2012}. Moreover, various types of topologically trivial Josephson junctions can also produce $4\pi$ periodic current-phase relations \cite{Kwon2004,Michelsen2008,Chiu2018,Zazunov2018b}.  

Fortunately, by investigating only slightly more elaborate devices, 
experiments could be in a position to detect very clear MBS signals that are much harder 
to fake.  For instance, in mesoscopic TS devices characterized by a strong Coulomb charging energy, highly nonlocal conductance phenomena are predicted for very low temperatures in the presence of zero-energy MBSs \cite{Fu2010,Beri2012,Altland2013,Beri2013}.   On the other hand, transport in a three-terminal device composed of a TS wire and two normal wires should yield characteristic MBS features in the current-current cross correlations between the normal wires  \cite{Haim2015,Haim2015b,Liu2015,Tripathi2016,Jonckheere2017,Zazunov2017,Zazunov2018a}:  While shot noise in two-terminal setups also carries interesting information \cite{Bolech2007,Nilsson2008,Golub2011,Wu2012,Liu2013,Giuliano2018}, in the three-terminal case already
its sign has an unconventional voltage dependence given by $ -{\rm sgn}(V_1V_2)$, where voltages $V_{1}$ and $V_2$ are applied between the TS and the respective normal wire \cite{Haim2015,Haim2015b,Liu2015,Tripathi2016,Jonckheere2017}.  A different --- and even more distinct --- Majorana manifestation in shot noise properties of topological trijunctions is described below.

\begin{figure} 
 \includegraphics[width=9.cm]{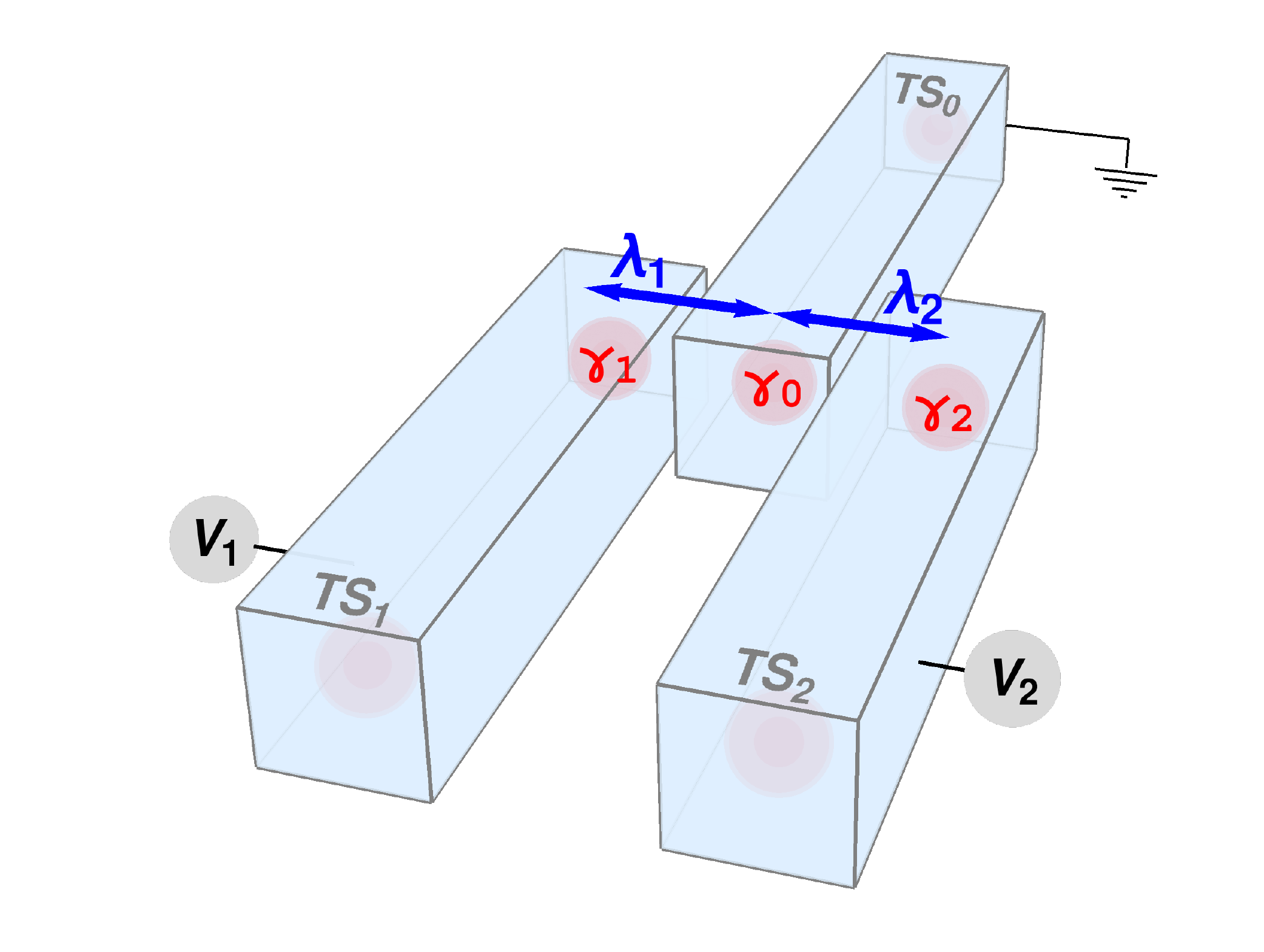}
\caption{Junction of three TS wires. The central wire (TS$_0$) with Majorana operator $\gamma_0$ is tunnel-coupled with amplitude $\lambda_{1}$ ($\lambda_2$) to the left, TS$_1$ (right, TS$_2$) wire with corresponding Majorana operator $\gamma_1$ ($\gamma_2$).  
A voltage $V_1$ ($V_2$) is applied between TS$_{1}$ (TS$_2$) and TS$_0$.  MBSs at the far ends are also indicated.}
\label{fig1} 
\end{figure}

We here point out that an experimentally identifiable and quite dramatic consequence of zero-energy MBSs arises when probing shot noise in  
a trijunction of three TS wires, see Fig.~\ref{fig1} for a schematic sketch. In this setup,  an unpaired zero-energy MBS must exist on general grounds \cite{Alicea2011}.  We show below that this MBS is 
directly responsible for giant shot noise levels.  We here define the shot noise amplitude from the current-current correlations measured in the left or right (TS$_{1}$, TS$_2$) wires in Fig.~\ref{fig1}, which are biased at voltages $V_{1}$ and  $V_2$ against the central (TS$_0$) wire, respectively.  The precise values of $V_{1}$ and $V_2$ are not crucial, and giant noise levels are found
at least for all commensurate cases, $p V_1 = q V_2$ with integer $p,q$~\cite{SM}.
(The case of non-commensurate voltages is more complex and cannot be accessed with the methods used below.)
We provide an intuitive explanation for the mechanism behind the giant noise levels by studying the atomic limit, where the TS gap $\Delta$ represents the largest energy scale. Calculations then simplify substantially and allow for an analytical understanding. 
By including above-gap continuum quasi-particles,
we next show that 
the shot noise amplitude 
is limited by a current-induced dephasing rate due to 
multiple Andreev reflection (MAR) processes.  The noise features are most pronounced at low voltage and small contact transparency, where the subgap current, and hence also the dephasing rate, is small. While the current shows similar MAR features as in TS-TS junctions \cite{Badiane2011,Houzet2013,Zazunov2016}, our results suggest that shot noise experiments for the setup in Fig.~\ref{fig1}
should readily find clear MBS signatures.

\emph{Model.---}The system is modeled by a generic  low-energy Hamiltonian, $H= \sum_{\nu=0,1,2} H_{{\rm TS}{_\nu}} + H_{t}$, 
where each TS wire corresponds to (we often put $e=\hbar=v_F=1$) \cite{Alicea2012}
\begin{equation}\label{eq:HTS}
H_{{\rm TS}{_\nu}} = \int_0^{\infty} dx \, \Psi^{\dagger}_{\nu}(x) \left( -i \partial_x \sigma_z
 + \Delta \sigma_y \right) \Psi^{}_{\nu}(x),
\end{equation}
with Nambu spinors $\Psi_{\nu}= (c^{}_{R,\nu}, c^{\dagger}_{L,\nu})^T$ and assuming chemical potential $\mu=0$.
Here $c_{L/R,\nu}$ are 
left/right-moving, effectively spinless fermion operators in the TS$_\nu$ wire, and Pauli matrices $\sigma_{x,y,z}$ (identity $\sigma_0$) act in Nambu space.  For notational simplicity, the gap $\Delta$ is assumed real and identical for all wires. 
The boundaries of the three wires at $x=0$ are connected by the tunneling Hamiltonian $H_t$.
With applied voltages $V_{j=1,2}$,  gauge-invariant phase differences are given by $\varphi_j(t) =2V_j t+\varphi_j(0)$.  We put $\varphi_j(0)=0$ but constant phase offsets could take into account, e.g., initial conditions or tunneling phase shifts.
We choose a gauge where the $\varphi_j(t)$ appear only in $H_t$ \cite{Zazunov2016,Jonckheere2017},
\begin{equation}\label{htun}
H_t= \sum_{j=1,2} \lambda_j \left(e^{i \varphi_j(t)/2 } c^{\dagger}_j c^{}_0 + {\rm h.c.} \right),
\end{equation}
with $c_\nu = [c_{L,\nu} + c_{R,\nu}](x=0)$.  In our units,  $\lambda_j$
 are dimensionless real tunneling amplitudes,  
 \begin{equation}\label{majdef}
 \lambda_1=\lambda \cos\chi ,\quad \lambda_2=\lambda \sin\chi,\quad 0 \leq \lambda \leq 1 ,
\end{equation}
  and the normal-state total transmission probability (`transparency') between TS$_0$ and TS$_1$, TS$_2$ is \cite{Jonckheere2017}
\begin{equation}\label{transparency}
\tau=\frac{4 \lambda^2}{(1+  \lambda^2)^2}.
\end{equation}

\emph{Keldysh approach.---}We solve this problem by using the Keldysh boundary Green's function (bGF) formalism \cite{Zazunov2016,Jonckheere2017}. The Keldysh bGF  of the uncoupled TS$_\nu$ wire is given by
$\check{g}_{\nu}(t - t')= -i \left \langle {\cal T}_C \Psi^{}_\nu(t) \Psi^{\dagger}_\nu(t') \right \rangle$,
with the boundary Nambu spinor $\Psi_\nu=(c^{}_\nu, c^{\dagger}_\nu)^T$ and the Keldysh time ordering operator ${\cal T}_C$. 
Retarded/advanced components of $\check{g}_\nu$ follow in frequency representation as \cite{Zazunov2016} 
\begin{equation}\label{bGFR}
g^{R/A}_\nu(\omega) =  \frac{\sqrt{\Delta^2 - (\omega \pm i 0^+)^2} \, \sigma_0  + \Delta \sigma_x}{\omega \pm i 0^+} .
\end{equation}
The $\omega=0$ pole in Eq.~\eqref{bGFR} describes the zero-energy MBS. Continuum quasi-particles appear at $|\omega|>\Delta$,
with boundary density of states $\sim \sqrt{\omega^2-\Delta^2}/|\omega|$ \cite{Zazunov2016}.
Physical quantities are expressed in terms of the full Keldysh bGF, $\check{G}$, 
which in turn follows by solving the Dyson equation,
$\check{G} = (\check{g}^{-1} - \check{W})^{-1}$,
where $\check{g}={\rm diag}_L(\check{g}_0,\check{g}_1,\check{g}_2)$ is diagonal in lead space. 
The tunneling matrix, $\check W={\rm diag}_K(W,-W)$, is diagonal in Keldysh space, where Eq.~\eqref{htun} yields the nonvanishing entries
\begin{equation}\label{htun2}
W_{0,j=1,2}(t) = \lambda_j \sigma_z e^{i\sigma_z\varphi_j(t)/2}, \quad W_{j,0}(t) = W_{0,j}^\dagger(t) .
\end{equation}
The time-dependent current flowing through TS$_j$, oriented toward the junction, corresponds to the Heisenberg operator
\begin{equation}\label{currentheis}
\hat I_j(t) = 2\frac{\partial H(t)}{\partial \varphi_j(t)} = i \Psi_j^\dagger(t) \sigma_z W_{j,0}(t) \Psi_0(t).
\end{equation}
With the average current  $I_j(t)=\langle \hat I_j(t)\rangle$, current-current correlations for
the TS$_{j=1,2}$ wires are defined as
\begin{equation} \label{fullnoise}
S_{jj'}(t,t') =\left \langle \hat I_j(t) \hat I_{j'}(t')\right \rangle - I_j(t) I_{j'}(t').
\end{equation}
Below we discuss the zero-frequency noise, $S_{jj'}\equiv S_{jj'}(\omega=0)$. For clarity,  we 
focus on the case $V_1=-V_2=V$ from now on (but see \cite{SM}). However, the atomic limit results below are identical for $V_1=V_2=V$.

\begin{figure}
\centering
\includegraphics[width=8.5cm]{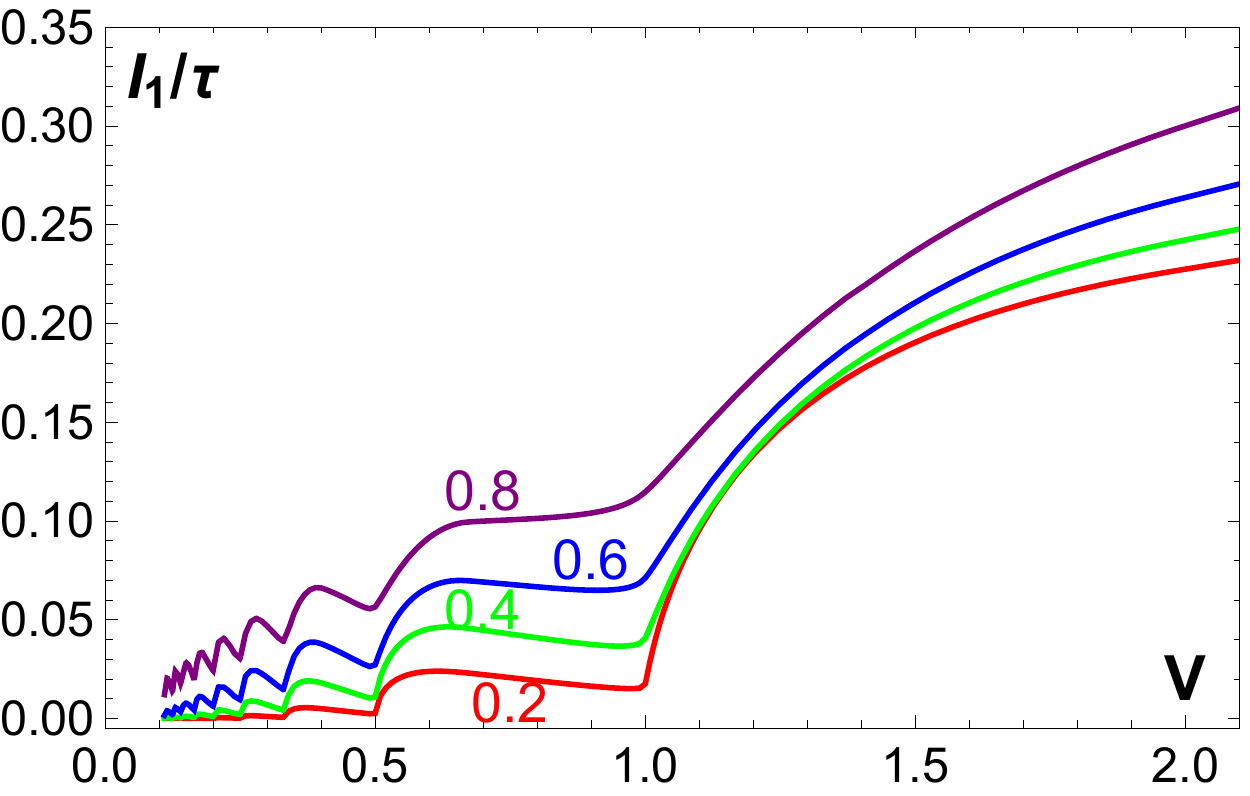}
\caption{Numerical results for the current $I_1$ (in units of $e \Delta/h$) vs voltage $V$ (in $\Delta/e$) for different transparencies $\tau$, see
Eq.~\eqref{transparency}, in a symmetric junction ($\lambda_1=\lambda_2$) with  $V_1=-V_2=V$.
For better visibility, $I_1$ is divided by $\tau$.  }
\label{fig2} 
\end{figure}

\begin{figure}
\centering
\includegraphics[width=8.5cm]{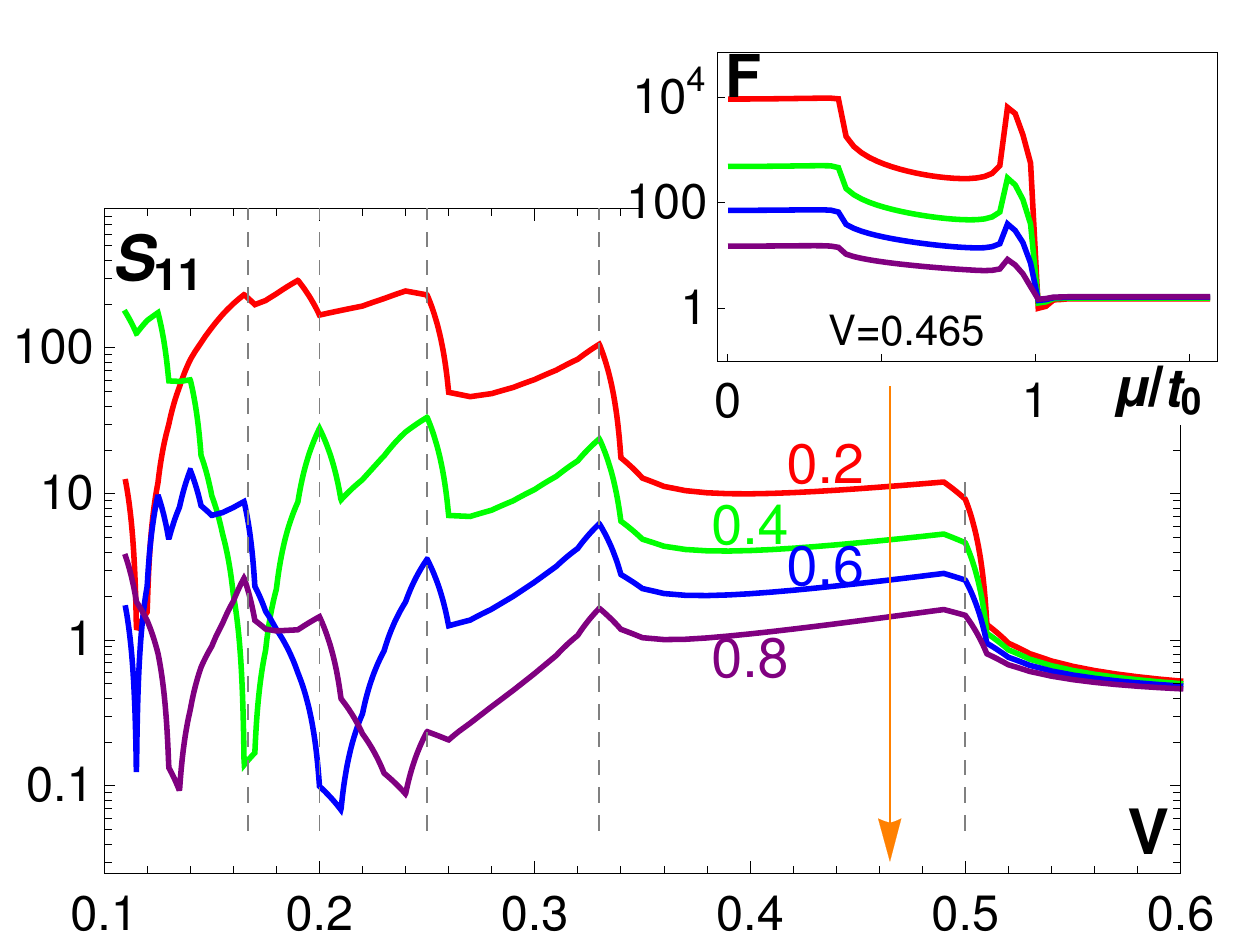}
\caption{Numerical results for shot noise $S_{11}$ (in units of $e^2 \Delta/h$) vs voltage $V$ (in $\Delta/e$) for different transparencies $\tau$ in a symmetric trijunction, cf.~Fig.~\ref{fig2}.  Dashed vertical lines mark MAR onsets, $V=e\Delta/n$ with $n=2,3,\ldots,6$.
Inset: Fano factor $F$ (on logarithmic scale) vs chemical potential $\mu/t_0$ of TS$_2$, for $V=0.465$ and bandwidth $t_0=10\Delta$. }
\label{fig3} 
\end{figure}

\emph{Numerical results.---}After a double Fourier transform along with a
summation over discrete frequency domains of width $V$, the Dyson equation  reduces to a matrix inversion problem which we have solved numerically, cf.~Ref.~\cite{Zazunov2006}. Given the solution for $\check{G}$, we  directly obtain the current-voltage characteristics as well as the zero-frequency shot noise amplitude.
Figure~\ref{fig2} shows numerical results for the current-voltage characteristics, with qualitatively similar features as for TS-TS junctions \cite{Zazunov2016,Badiane2011,Houzet2013}. 
In particular,  MAR onsets are visible at $V=\Delta/n$ (integer $n$), and 
for low transparency and small $V$, the current becomes very small.
 Figure~\ref{fig3} illustrates our numerical shot noise results for $S_{11}(V)$.
 In contrast with the current, shot noise behaves in a totally different manner as compared to TS-TS junctions \cite{Aguado2017,Houzet2013}. 
 Taking note of the logarithmic noise scale in Fig.~\ref{fig3}, we observe giant noise levels which are particularly pronounced near  MAR onsets. Remarkably, in contrast to the average current, the noise amplitude shows an overall \emph{increase} when reducing the transparency $\tau$.  
 The inset of Fig.~\ref{fig3} demonstrates that these features are directly related to MBSs: The Fano factor, $F=S_{11}/(2eI_1)$, becomes small when one lead (here TS$_2$) exits the topological regime $|\mu|/t_0<1$ upon changing its chemical potential $\mu$ (with $\mu=0$ in the other wires). Using the bGFs in Ref.~\cite{Jonckheere2017}, 
 we find very large $F$ for all $|\mu|/t_0<1$ (especially at small $\tau$), with an abrupt drop down to $F\simeq 1$ 
 for $|\mu|/t_0>1$. We next show analytically that the giant noise levels are tied
 to the existence of an unpaired zero-energy MBS.

\emph{Atomic limit.---}Since the features in Fig.~\ref{fig3} are most pronounced for small $V$ and low transparency,
we consider the atomic limit where $\Delta$ represents the largest energy scale and the bGF \eqref{bGFR} simplifies to  
\begin{equation}\label{bGFR1}
    g^{R/A}_\nu(\omega)=\frac{\Delta}{\omega\pm i\eta} \left( \begin{array}{cc} 1 & 1 \\ 1 & 1 \end{array} \right).
\end{equation}
 The small parameter $\eta>0$ represents a finite parity relaxation rate (see below).  
 By construction, the simplified bGF \eqref{bGFR1}  neglects above-gap continuum states. Boundary fermions are 
 thus projected to the Majorana sector, $c_{\nu} \to \sqrt{\Delta}  \, \gamma_{\nu}$,
where Majorana operators, $\gamma_{\nu} = \gamma_{\nu}^{\dagger}$, satisfy the anticommutation relations
$\{ \gamma_{\nu}, \gamma_{\nu'}\} = \delta_{\nu\nu'}$. The atomic limit Hamiltonian for an arbitrary trijunction thereby follows from the full $H(t)$ as, see Eq.~\eqref{majdef},
\begin{eqnarray} \label{hat}
H_{\rm at}(t) &=&  2i\Omega(t) \left[\cos(\chi)\gamma_1-\sin(\chi)\gamma_2\right] \gamma_0 ,\\ \nonumber \Omega(t) &=& \lambda \Delta \sin\left(Vt\right).
\end{eqnarray}
By passing to a rotated Majorana basis,
\begin{eqnarray}
\gamma_-&=&\cos(\chi)\gamma_1-\sin(\chi)\gamma_2,
\\ \nonumber
\gamma_+&=&\sin(\chi)\gamma_1+\cos(\chi)\gamma_2,
\end{eqnarray}
and combining $\gamma_-$ and $\gamma_0$ to a complex fermion, 
    $d=(\gamma_- + i \gamma_0)/\sqrt2$, one can solve the problem in an elementary manner.
Indeed,  $i\gamma_-\gamma_0=d^\dagger d-1/2$ is the only combination of Majorana operators appearing in $H_{\rm at}$,
and Eq.~\eqref{hat} thus affords the alternative representation
\begin{equation}\label{eq:Hd}
H_{\rm at}(t) = 2i\Omega(t)\gamma_-\gamma_0= \Omega(t) (2d^{\dagger} d - 1) ,
\end{equation}
where the parity $(-1)^{d^\dagger d}$ is always conserved.  
The Majorana operator $\gamma_+$, on the other hand, does not show up in the Hamiltonian and represents the zero-energy MBS 
of the trijunction.  
Expressing $\gamma_+=(f+f^\dagger)/\sqrt2$ in terms of 
a  zero-energy fermion $f$,
the current operator \eqref{currentheis} takes the form (say, for TS$_1$) 
\begin{eqnarray}\nonumber 
\hat I_1(t) &=& 2i\lambda_1\Delta\cos(Vt) \left[
\cos( \chi) \gamma_- + \sin(\chi) \gamma_+ \right]\gamma_0 
\\\label{curr2} &=&\lambda_1\Delta\cos(Vt)
\Bigl [
\cos\chi \ ( 2d^{\dagger} d - 1) \\ \nonumber &&\quad + \sin\chi \ 
  ( f+f^\dagger ) (d- d^{\dagger})\Bigr].
\end{eqnarray}
The non-trivial coupling between the $d$ fermion and the zero-mode fermion $f$  
 in Eq.~\eqref{curr2} is ultimately responsible for giant noise levels.  Although $f$ does not appear in 
the Hamiltonian, it affects the current operator when all three TS wires are coupled together.

\begin{figure}
\centering  
\includegraphics[width=8cm]{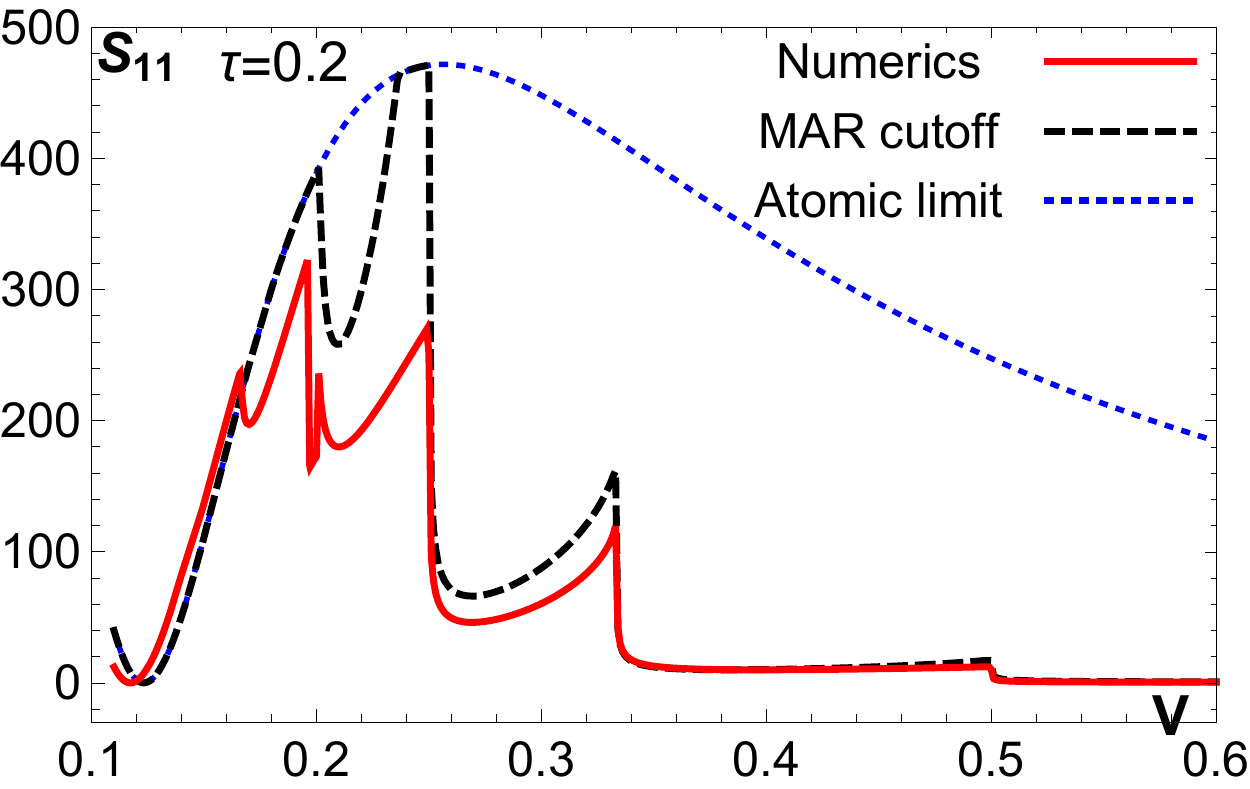}\\ 
\vspace{0.1cm}
\includegraphics[width=8cm]{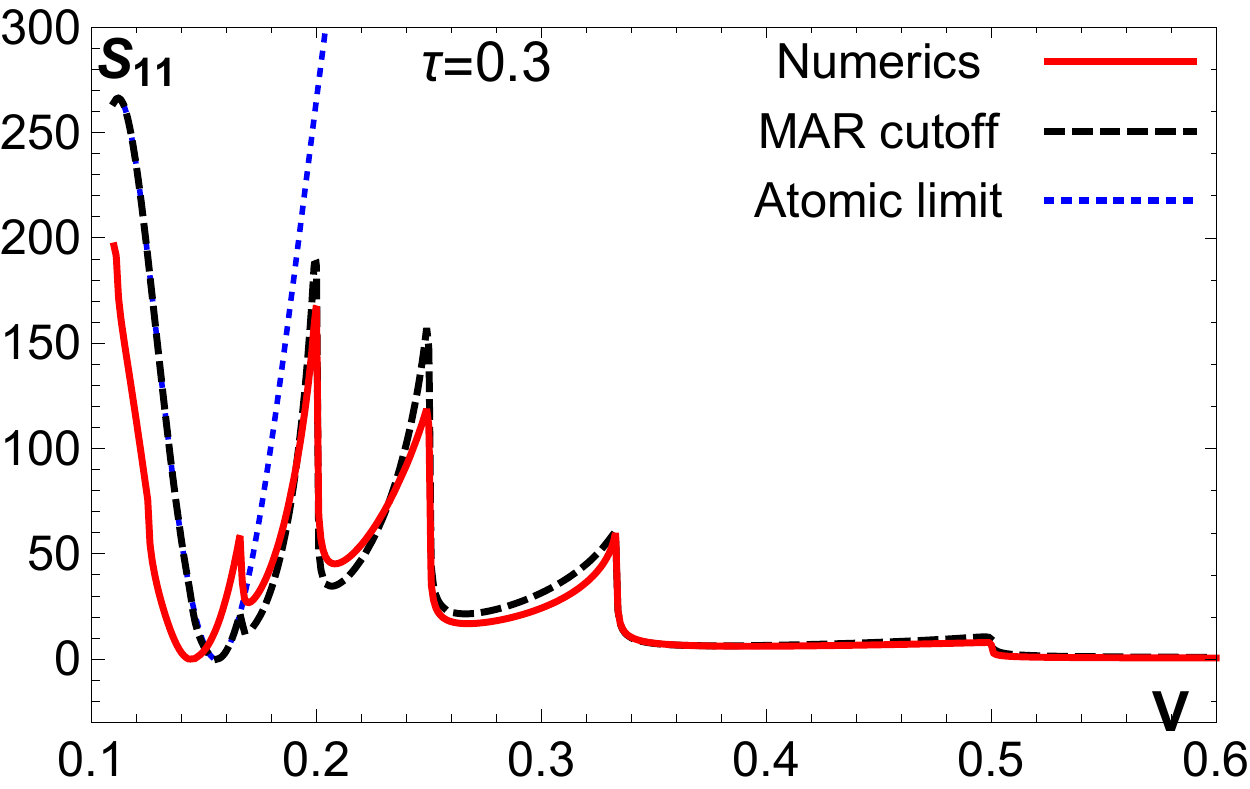}
\caption{Shot noise $S_{11}$ vs voltage $V$ for $\tau=0.2$ (top) and $\tau=0.3$ (bottom) in a symmetric trijunction.  The atomic limit prediction \eqref{eq:S11atomic} is shown
for $\eta=10^{-5}\Delta$ as blue dotted curve, full numerical results as solid red curves. Black dashed curves include MAR effects, see Eqs.~\eqref{eq:S11atomic} and \eqref{eq:S11MAR}.}
\label{fig4}
\end{figure}

In $(d,f)$ fermion representation,  physical steady state density matrices must commute with $H_{\rm at}$ and therefore have
the form $\rho_{\rm s}= \sum_{n,m=0,1} w_{nm} |nm\rangle  \langle nm|$, 
where $w_{nm}\ge 0$ with $\sum_{nm} w_{nm}=1$
is the statistical weight of the state $|nm \rangle=(d^\dagger)^n (f^\dagger)^m|0\rangle$.  For a symmetric trijunction, we then obtain 
the average current in the atomic limit as
\begin{equation}\label{eq:mI1atomic}
I^{({\rm at})}_1(t) = \lambda \Delta \cos(V t) (\langle d^\dagger d\rangle - 1/2).
\end{equation}
As for a TS-TS junction \cite{Aguado2017}, only the AC current with frequency $\omega=V$ can be finite. 
For the shot noise, with Eq.~\eqref{curr2} and the Bessel $J_1$ function, we find \cite{SM}
\begin{equation}\label{eq:S11atomic}
S^{({\rm at})}_{11} = \frac{\lambda^2 \Delta^2}{ 4\eta}  J_1^2\left(2 \lambda \Delta/V \right) ,
\end{equation}
which is limited only by the parity relaxation time $1/\eta$.  Examples for 
Eq.~\eqref{eq:S11atomic} are shown in Fig.~\ref{fig4} and, for small $V$, agree rather well with the full numerics.  
For larger $V$, the complex peak structure in $S_{11}(V)$ is missed by Eq.~\eqref{eq:S11atomic} and the noise level is overestimated. 
Figure \ref{fig4} also shows a marked noise minimum at low voltage  which shifts to smaller $V$ as $\tau$ decreases. The position of the minimum corresponds to    the first zero of the Bessel function in Eq.~(\ref{eq:S11atomic}).
Similar noise dips are also observable in the full numerical results in Fig.~\ref{fig3}.  

\emph{Discussion.---}The giant noise features are deeply related to the existence of the
zero mode $\gamma_+$, which also implies that the current operator and the Hamiltonian  do not  commute.  One can understand the giant noise as a generic feature of periodically driven two-level systems.
To that end, we note that three Majorana operators, $\gamma_{0,1,2}$, can  equivalently be represented in terms of Pauli matrices. Choosing
\begin{equation}
    \tau_z= 2i\gamma_1\gamma_0, \quad \tau_x= 
    2i\gamma_0\gamma_2, 
\end{equation}
we obtain the current operator,
Eq.~\eqref{curr2}, in diagonal form,
$\hat I_1(t)= \lambda_1\Delta\cos(Vt) \tau_z$.
However, in this basis, 
$H_{\rm at}(t) =\Omega(t)[\cos(\chi)\tau_z + \sin(\chi)\tau_x]$ is not diagonal anymore.
 Since the $\tau_x$ part in $H_{\rm at}$  
coherently rotates $\tau_z$ and hence $\hat I_1(t)$,  we directly encounter a  coherent current switch which has divergent shot noise in the absence of relaxation channels. 
Moreover, since a zero-energy MBS always exists in a TS trijunction \cite{Alicea2011}, the giant noise
features are robust when adding a finite hybridization between $\gamma_1$ and $\gamma_2$.

A complementary viewpoint follows by noting that
the uncoupled system has three MBSs at the junction, where $\gamma_0$ resides at energy $E=0$ while $\gamma_1$ ($\gamma_2$) correspond to $E=V$ ($E=-V$).  Including the tunnel couplings,
a resonant process similar to crossed Andreev reflection (CAR) exists 
where two electrons are emitted from TS$_0$. One of them enters TS$_1$ through  $\gamma_1$, the other TS$_2$ via $\gamma_2$. In a sequential tunneling picture, the rate for this process is 
\begin{equation}\label{Gamma}
\Gamma = \lambda^2 \Delta^2\int  dE
 \left(\frac{\eta}{(E-V)^2+\eta^2}\right)^2 
 \frac{\lambda^2\Delta^2}{E^2 + \eta^2}.
\end{equation}
The first factor in the integrand comes from the density of states
for the MBSs $\gamma_1$ and $\gamma_2$, while the second is due to the probability for a CAR process. To leading order in $1/\eta$, Eq.~\eqref{Gamma} yields
$\Gamma =\lambda^4 \Delta^4/ (4\eta V^2)$. The sequential tunneling result for $S_{11}$ 
then coincides with Eq.~(\ref{eq:S11atomic}) to lowest order in $\lambda\Delta/V$ \cite{SM}.
We remark that in fully transparent S-S junctions,  thermal noise  exhibits a similar phenomenon
\cite{Alvaro1996,Averin1996}.
Since MBSs are equal-probability superpositions of electrons and holes, 
 the corresponding hole process also exists.
We thus encounter no average DC current yet have giant shot noise.

\emph{MAR effects.---}Finally, we take into account continuum states \cite{SM}. To that end, we split the boundary fermion
 as $c_{\nu} = \sqrt{\Delta} \gamma_{\nu} + a_{\nu}$, with the Majorana part as before but now supplemented by above-gap fermions ($a_{\nu}$).
$H_t$ then includes (i) MBS-MBS couplings as in Eq.~\eqref{hat}, (ii)  MBS-continuum couplings, and (iii) continuum-continuum terms. The latter terms 
are irrelevant for $V\ll \Delta$ and low transparency, while type (ii) terms, which correspond to MAR processes, 
can
change the parity $(-1)^{d^\dagger d}$. This implies a loss of coherence for the $d$ fermion dynamics. The average time between two tunneling processes of type (ii) defines a long-time cutoff, $T_{\rm MAR}$, limiting the integration of current correlations.
A good approximation is given by
$T_{\rm MAR}(V)= N_1/I_1(V)$,
where $N_1=1+ \lfloor \Delta/V \rfloor$ is the number of electrons transferred in one MAR process.  
The dominant MAR effects on shot noise can then be taken into account by replacing $\eta$ in Eq.~\eqref{eq:S11atomic} by a voltage-dependent effective parity relaxation rate,
\begin{equation}
\eta\to\eta_{\rm eff}(V) = {\rm max}\left( T^{-1}_{\rm MAR}(V), \eta\right),
\label{eq:S11MAR}
\end{equation}
where $\eta$ is here due to additional parity relaxation channels and `parity' refers to the Majorana sector only.
Results obtained from Eq.~\eqref{eq:S11MAR} are shown in Fig.~\ref{fig4} and 
exhibit  quantitative agreement with our full numerics. In particular, the peak pattern is now correctly reproduced without fitting parameter.  The agreement is not quantitative when $T_{\rm MAR}\approx 1/\eta$, where Eq.~\eqref{eq:S11MAR} is too simplistic, cf.~the case $\tau=0.2$ in Fig.~\ref{fig4}.

\emph{Conclusions.---}The topological trijunction in Fig.~\ref{fig1} provides an attractive setup for experimental studies: an unpaired zero-energy MBS is directly responsible for giant shot noise. Moreover, by measuring the detailed voltage dependence of the shot noise, precious information on parity relaxation rates can be obtained.  If the MBSs are tunnel-coupled to additional low-energy states, e.g., because of finite wire length or due to fermion states localized near the junction, we expect a partial suppression of the shot noise amplitudes \cite{SM}.   However, extrinsic noise sources are at odds with the predicted MAR features and can easily be ruled out.  Finally, let us note that similar giant shot noise might be obtained in systems containing more than 3 TS electrodes - in particular for an odd number of TS (e.g. 5) one expects that a zero-mode should always be present. However the strong robustness with respect to the parameters might be specific to the 3TS case, which is also the most accessible experimentally.

\begin{acknowledgments}
This work has been supported by the Excellence Initiative of Aix-Marseille University -- A$\ast$MIDEX, a French `investissements d'avenir' program, 
 by the Deutsche Forschungsgemeinschaft within Grants No.~EG 96/11-1 and CRC TR 183 (project C04), by the Spanish
MINECO through Grant Nos.~FIS2014-55486-P and FIS2017-84860-R, and through the `Mar{\'i}a de Maeztu' Program 
(MDM-2014-0377).
\end{acknowledgments}

\appendix

\section{Shot noise in the atomic limit}
\label{sec1}
In this section, we outline the calculation of the zero-frequency noise in the TS$_1$ lead,
which is given by
\begin{equation}\label{sm1:S11}
S_{11}= \frac{1}{T_V}\int_0^{T_V} dt_+ \int_{-\infty}^\infty dt_-
S_{11}\left(t_+ +\frac{t_-}{2},t_+-\frac{t_-}{2}\right),
\end{equation}
where $T_V = 2\pi/V$. The current-current correlator $S_{11}(t,t')$ is defined by Eq.~(8) in the main text.
In the atomic limit, one has [cf. Eqs.~(10)-(14) in the main text]
\begin{equation}
S_{11}^{({\rm at})}(t,t') = {\rm tr}\left\{ \rho_{\rm s} \hat {\cal I}_1(t) \hat {\cal I}_1(t') \right\} -
I^{({\rm at})}_1(t) I^{({\rm at})}_1(t'),
\end{equation}
where $\hat {\cal I}_1(t) = {\cal U}^\dagger(t,0) \hat I_1(t) {\cal U}(t,0)$,
${\cal U}(t,0) = {\cal T} \exp\{ -i \int_0^t d\tau H_{\rm at}(\tau) \}$
is the time-evolution operator, ${\cal T}$ is the time-ordering operator, and
the (steady state) density matrix $\rho_{\rm s}$ has been introduced in the main text.
For a symmetric junction, $\lambda_1=\lambda_2=\lambda/\sqrt2$, one obtains
\begin{eqnarray}
S_{11}^{({\rm at})}(t,t') &=& \lambda^2 \Delta^2  \cos (V t)  \cos (V t') \\ \nonumber 
&\times& \left[
{\rm tr} \left\{ \rho_{\rm s} X(t) X(t') \right\} -  \left( n_d - {1\over2}\right)^2 \right],
\end{eqnarray}
where $n_d = {\rm tr} \left\{ \rho_{\rm s} d^\dagger d \right\}$ and
\begin{equation}
X(t) = d^\dagger d - \frac12 + \frac12 \left( f + f^\dagger \right) \left( d e^{-i \Phi(t)} - d^\dagger e^{i \Phi(t)} \right)
\end{equation}
with $\Phi(t) = (2\lambda\Delta/V) [ 1 - \cos(Vt) ]$.
Taking the trace over the fermions $(d, f)$ yields
\begin{eqnarray}\label{sm1:S11xx}
&& S_{11}^{({\rm at})}(t,t') =  \lambda^2 \Delta^2 \cos (V t)  \cos (V t') 
\Biggl[ n_d \left( 1 - n_d \right) \nonumber \\ 
&&\quad + \frac{1}{4}
\left\{ \cos \Phi(t, t') + i \left( 2 n_d - 1 \right) \sin \Phi(t, t') \right\} \Biggr],
\end{eqnarray}
where $\Phi(t,t') = \Phi(t)-\Phi(t')$.
Substituting Eq.~\eqref{sm1:S11xx} into Eq.~\eqref{sm1:S11} and using the expansion \cite{GR}
\begin{equation}\label{sm1:Bess}
e^{i z \cos \theta} = \sum_{k=-\infty}^\infty i^k J_k(z) e^{i k \theta},
\end{equation}
where $J_k(z)$ is the Bessel function of order $k$,
one arrives at a formally divergent expression for the zero-frequency noise,
\begin{equation}\label{sm1:S11x}
S_{11}^{({\rm at})} = \frac{\pi}{2}
\delta(\omega = 0) \lambda^2 \Delta^2 J_1^2 \left( 2\lambda\Delta/V \right).
\end{equation}
Note that in contrast to the current $I^{({\rm at})}_1(t)$, the noise $S_{11}^{({\rm at})}$ does
not depend on the state of the Majorana fermion subsystem.
Equation~\eqref{sm1:S11x} is then regularized by introducing a finite parity relaxation rate $\eta$,
cf. Eq.~(9) in the main text, with $2 \pi \delta(0) = \int dt_- \rightarrow 1/\eta$.
Hence we obtain Eq.~(15).

Using the asymptotic forms of $J_1(z)$ at small and large $z$,
one gets, respectively,
\begin{equation}
S_{11}^{({\rm at})} \sim \frac{\lambda^4 \Delta^4}{4\eta V^2}~~~\textrm{for $\lambda \Delta \ll V$},
\end{equation}
\begin{equation}
S_{11}^{({\rm at})} \sim \frac{\lambda \Delta V}{ 4 \pi \eta}
\cos^2\left({2 \lambda \Delta\over V} - {3\pi\over 4} \right)~~~\textrm{for $\lambda \Delta \gg V$}.
\end{equation}
The crossover from $1/V^2$ to linear in $V$ behavior with decreasing $V$
indicates that $S_{11}^{({\rm at})}$ must vanish in the limit $V \rightarrow 0$ (not accessible numerically).
At the same time, $S_{11}^{({\rm at})}$ exhibits oscillations with $1/V$ at sufficiently low $V$.

\section{Dissipative effects on Majorana-induced noise}

At low voltage $|V| \ll \Delta$, MAR processes may trigger
transitions between the adiabatic Andreev level ($d$-fermion) and continuum states above the gap.
These processes cause random flips of the pseudospin $S_d = d^\dagger d - 1/2$,
which eventually leads to a suppression of the supercurrent noise associated with the zero mode $f$.

\begin{figure}[t!]
\centering
\includegraphics[width=8.5cm]{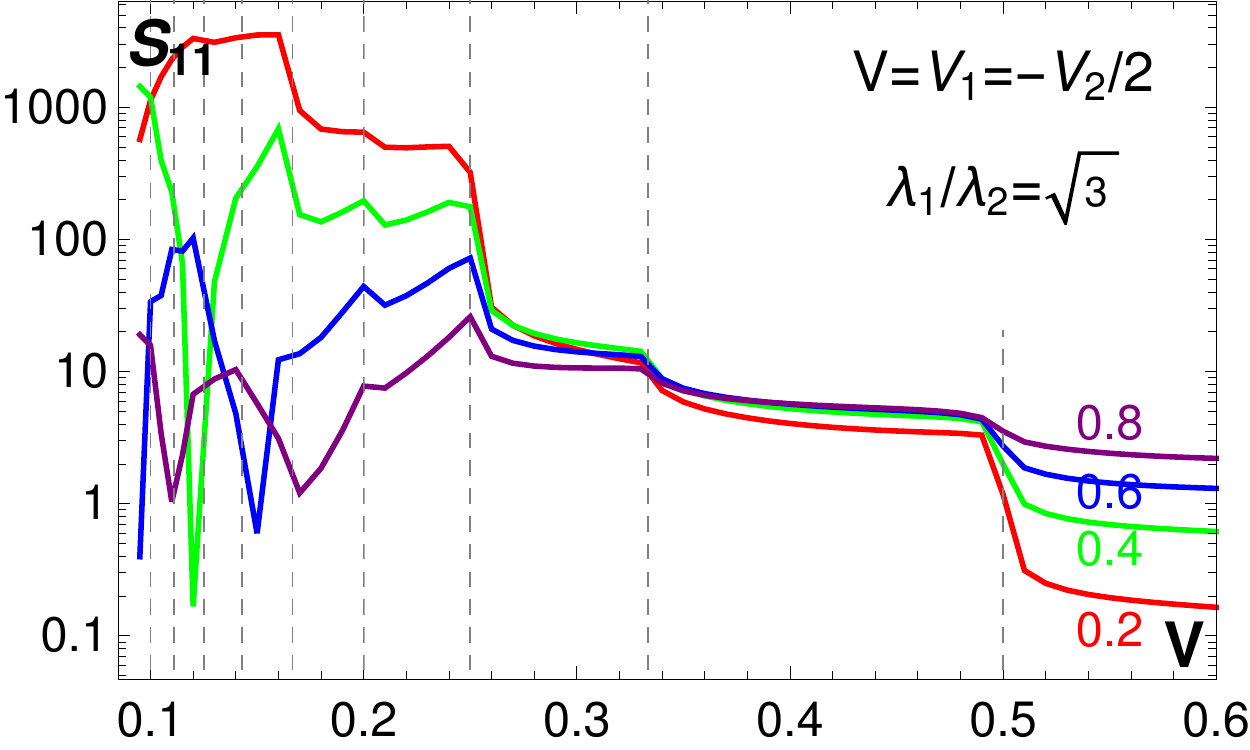}

\vspace*{0.5cm}
\includegraphics[width=8.5cm]{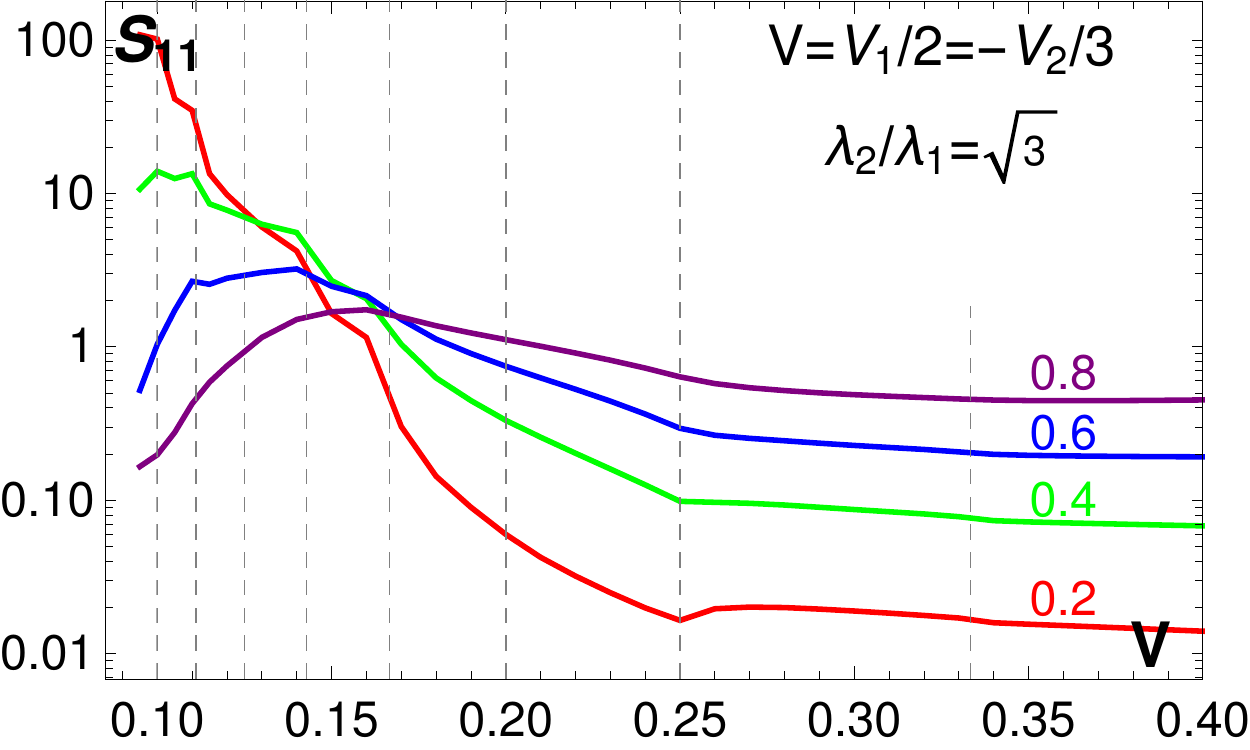}
\caption{
Shot noise $S_{11}$ (in units of $e^2\Delta/h$) vs voltage $V$ (in $\Delta/e$) for different transparencies from $\tau=0.2$ to $\tau=0.8$. We study two commensurate voltage configurations, $pV_1=qV_2$. The top panel is for $p=2$ and $q=-1$, with asymmetric tunnel couplings $\lambda_1/\lambda_2=\sqrt{3}$. The bottom panel shows the case $p=3$ and $q=-2$ for $\lambda_1/\lambda_2=1/\sqrt{3}$. } 
\label{fSM} 
\end{figure}

Assuming that (i) the noise due to pseudospin fluctuations is dominating and that
(ii) hybridization of the $f$-fermion with continuum states is very weak and can be neglected,
one obtains
\begin{eqnarray}\label{sm2:S11Sd}
&&
S_{11}(t, t') \approx  \lambda^2 \Delta^2 \cos(Vt) \cos(Vt') \times \\ \nonumber &&
\left[\left\langle
\delta S_d(t) \delta S_d(t')\right \rangle + {1 \over 4}
\Big\langle d^\dagger(t)d(t') + d(t)d^\dagger(t') \Big\rangle \right],
\end{eqnarray}
with $\delta S_d(t) = S_d(t) - \langle S_d(t) \rangle$.
On long time scales, $|t-t'| \gg \omega_S^{-1}$,
where $\omega_S$ is the frequency of pseudospin flips
due to quasiparticle tunneling, one has
$\langle \delta S_d(t) \delta S_d(t') \rangle \approx 0$, with $\langle S_d(t) \rangle = 0$.
For short time scales, $|t-t'| \lesssim \omega_S^{-1}$, we still have
\begin{equation}
\Big\langle d^\dagger(t)d(t') + d(t)d^\dagger(t') \Big\rangle \approx
{1 \over 2} \, e^{i \Phi(t, t')} + {\rm c.c.}
\end{equation}
Applying again the identity \eqref{sm1:Bess} and averaging over the 'center-of-mass' time $t_+$,
cf. Eq.~\eqref{sm1:S11}, the zero-frequency shot noise takes the  form
\begin{eqnarray}\label{sm2:S11cos}
S_{11} &=& \frac{\lambda^2 \Delta^2}{16} \int_{-\infty}^\infty d \tau \, \sum_{s,s' = \pm 1} i^{s+s'}
 \nonumber \\ &\times&
\sum_{n = -\infty}^\infty J_{n+s}(Q/V) J_{n-s'}(Q/V) \cos(n V \tau),
\end{eqnarray}
with $Q=2\lambda \Delta$. Here the $nV$ harmonics are associated with the charge transfer $ne$ due to MAR processes.
Pseudospin flips are now readily incorporated by replacing
$e^{i n V \tau} \rightarrow e^{i n V \tau - \Gamma_n |\tau|}$ in Eq.~\eqref{sm2:S11cos},
where the partial rates $\Gamma_{n \geq 1} = \Gamma_{-n}$ can be estimated similarly as
in a two-terminal case, cf. Ref.~\cite{ALY}, while $\Gamma_0 \sim \eta$ is the parity relaxation rate 
 in the absence of MAR processes.
As a result, we obtain
\begin{eqnarray}\label{sm2:S11}
S_{11} &=& \frac{\lambda^2 \Delta^2}{8} \sum_{n = -\infty}^\infty {\Gamma_n \over n^2 V^2 + \Gamma_n^2}\\ \nonumber &\times&
\left[ J_{n+1}(Q/V) - J_{n-1}(Q/V) \right]^2.
\end{eqnarray}
In particular, in the atomic limit one has $\Gamma_n = \Gamma_0 \delta_{n0}$,
and Eq.~\eqref{sm2:S11} reduces to Eq.~(15) in the main text.

Suppression of the giant noise can also arise
from additional subgap states hybridized with the Majorana fermions $\gamma_\nu$ at the trijunction.
For instance, this can be due to (i) exponentially small couplings between
Majorana states located at opposite ends of finite-length TS wires and/or due to 
(ii) hybridization between $\gamma_\nu$ and low-energy impurity states localized near the contact region.
At the phenomenological level, such `quasiparticle poisoning' effects can be taken into account
by introducing a corresponding parity relaxation rate, $\Gamma_{\rm qp}$,
cf. Eq.~\eqref{sm2:S11Sd},
\begin{eqnarray}
S_{11}(t, t') \approx  \lambda^2 \Delta^2 \cos(Vt) \cos(Vt')
\Bigl [ \langle \delta S_d(t) \delta S_d(t') \rangle
\nonumber \\ 
+ {1 \over 4}
\left\langle d^\dagger(t)d(t') + d(t)d^\dagger(t') \right\rangle e^{- \Gamma_{\rm qp} |t -t'|} \Bigr].
\hspace{.5cm}
\end{eqnarray}
As a result, $\Gamma_{\rm qp}$ is added to the rates of MAR subharmonics,
implying $\Gamma_n \rightarrow \Gamma_n + \Gamma_{\rm qp}$ in Eq.~\eqref{sm2:S11}.

\section{Other configurations}

We here demonstrate that giant noise appears in general for commensurate voltage configurations, $pV_1=qV_2$ with integer $p,q$, where the case $pq=\pm 1$ has been studied in the main text.  In Fig.~\ref{fSM}, we show numerical results for two other examples, where we also allow for
asymmetric tunnel couplings, $\lambda_1/\lambda_2\neq 1$. The results in Fig.~\ref{fSM} illustrate that giant noise is generically observed for commensurate voltages.
We note that for larger values of $|pq|$, somewhat 
lower $V$ are required to reach comparably high noise levels. Finally,
Fig.~\ref{fSM}  also underlines the robustness of  giant noise features against asymmetries in the tunnel couplings.  In fact, this robustness already follows from our analytical calculations in the atomic limit, see Sec.~\ref{sec1} and the main text.


\begin{thebibliography}{99}
\bibitem{SM} See the appendix, where we provide additional details on the derivation of Eq.~\eqref{eq:S11atomic}, about the effects of MAR processes and of low-lying subgap quasi particles  on shot noise, and about other voltage configurations.
\end{thebibliography}
\end{document}